\documentclass[aps,prl,onecolumn,superscriptaddress,showpacs,12pt]{revtex4}

\usepackage{graphics}
\usepackage{bm}
\usepackage{amsmath}    
\usepackage{amssymb}    
\usepackage{graphicx}   
\usepackage[hang]{subfigure}

\begin{document}

\title {Planar Josephson Tunnel Junctions\\
 in an Asymmetric Magnetic Field}

\author{R.\ Monaco}
\affiliation{Istituto di Cibernetica del C.N.R., 80078, Pozzuoli,
Italy and Dipartimento di Fisica, Universita' di
Salerno, 84081 Baronissi, Italy.}\email
{roberto@sa.infn.it}

\date{\today}
\begin{abstract}
We analyze the consequences resulting from the
asymmetric boundary conditions imposed by a non-uniform external
magnetic field at the extremities of a planar Josephson tunnel
junction and predict a number of testable signatures. When the
junction length $L$ is smaller than its Josephson penetration depth
$\lambda_j$, static analytical calculations lead to a Fresnel-like
magnetic diffraction pattern, rather than a Fraunhofer-like one
typical of a uniform field. Numerical simulations allow to investigate
intermediate length ($L\approx \lambda_j$) and long ($L>\lambda_j$)
junctions. We consider both uniform and $\delta$-shaped bias
distributions. We also speculate on the possibility of exploiting
the unique static properties of this system for basic experiments
and devices.
\end{abstract}

\maketitle

\section{INTRODUCTION}

Both the static and dynamics properties of a Josephson tunnel
junction (JTJ) are affected by the presence of an externally applied
magnetic field. Since the discovery of the Josephson effect in 1962,
the magnetic diffraction phenomena of the supercurrent and the
occurrence of current singularities (Fiske steps) due to resonant
cavity modes have been studied under the assumption of an
homogeneous magnetic field\cite{barone}. The aim of this letter is to overcome this textbook assumption and to investigate the
consequences of enforcing asymmetric boundary conditions (ABC) on
rectangular JTJs having length $L$ along the $X$ axis, width $W$
along the $Y$ axis and uniform critical
current density $J_c$. We assume the junctions to be one-dimensional,
i.e., $W<<L,\lambda_j$, with $\lambda_{j}=\sqrt{\Phi_0/2\pi d_{e}
\mu _{0} J_{c}}$ being the Josephson penetration depth\cite{joseph}
(where $\Phi_0$ is the magnetic flux quantum, $d_e$ the effective
junction magnetic penetration\cite{wei} and $\mu_0$ the vacuum
permeability). It is well known that, under these circumstances, the
gauge-invariant phase difference $\phi$ of the order parameters of
the superconductors on each side of the tunnel barrier obeys the
static or d.c. perturbed sine-Gordon equation\cite{barone}:

\begin{equation}
 \sin \phi(X) = \lambda_j^2 \frac{d^2 \phi(X)}{d X^2}+ \frac{\mathcal{I}_b(X)}{J_c W} , \label{sG}
\end{equation}

\noindent in which the term $\mathcal{I}_b(X)$ is the distribution
of the externally applied bias current $I_b$; choosing the $X$-axis
origin in the center of the junction, $I_b=
\int^{L/2}_{-L/2}\mathcal{I}_b(X) dX$. In normalized units of
$x=X/\lambda_j$, Eq.(\ref{sG}) becomes:

\begin{equation}
\phi_{xx} = \sin \phi(x) - \gamma(x). \label{PDE}
\end{equation}

\noindent With such normalization the junction length is
$l=L/\lambda_j$; further, the Josephson (zero-voltage) current
$i_j$ through the barrier is obtained as the spatial average of
$\sin \phi$, $i_j=\left\langle \sin \phi
\right\rangle=(1/l)\int_{-l/2}^{l/2} \sin \phi(x) dx$, while the
total bias current $i_b=I_b/(J_c WL)$ is given by the spatial
average of $\gamma$, $i_b=\left\langle \gamma \right\rangle$. In
this work we will consider two quite different symmetric bias
current profiles: i) uniform bias $\gamma(x)=\gamma_u$ for which
$i_b=\gamma_u$ and ii) $\delta$-shaped bias $\gamma(x)=
\gamma_{\delta} \delta(x)$ for which, being
$\int_{-a}^{a}\delta(x)dx=1$, we have $i_b=\gamma_{\delta}/l$.
$\delta$-biased JTJs in a uniform magnetic field have been the
subject of a recent theoretical and experimental
investigation\cite{PRB10,PRB09}. According to the magnetic Josephson
equation\cite{joseph}, the boundary conditions for Eq.(\ref{PDE})
depend on the external field values at the junction extremities:

\begin{equation}
\phi_x(-l/2)=h_L \qquad \phi_x(l/2)=h_R , \label{bc}
\end{equation}

\noindent in which $h_{L,R}$ are the $Y$-components $H_{L,R}$ of the
applied magnetic field at the left and right junction ends,
normalized to $\Phi_0/(2 \pi \mu_0 d_e \lambda_j)$. Classically one
considers the case of a uniform field applied in the junction plane
perpendicular to the long junction dimension $L$, so that $h_R=h_L$.
We will refer to the above conditions as {\it symmetric} or
classical boundary conditions to distinguish them from the {\it
asymmetric} boundary conditions (ABC) achieved when the magnetic
fields at the junction extremities have the same amplitudes, but
opposite directions, that is, $h_R=-h_L\equiv h_{a}$. We will
analyze the properties of ABC later on; for now we only remark that
they reminds of the boundary conditions for an in-line symmetric
junction\cite{OS} with $h_{L,R}$ being the self-fields produced by
the external current which enters the junction at one extremity and
leaves at the opposite one. Figs.1(a) and (b) show two
examples of how the ABC could be realized in practical devices. In
both cases the asymmetry is obtained by the current $I_{cl}$ flowing
in a properly designed control line. The control line technique has
been widely and successfully used to produce local magnetic fields
in Josephson structures since 1994\cite{zhang}. In Fig.1(a)
the control line, separated from the junction by an insulating
layer, runs aside the long dimension of an overlap-type planar JTJ
and flips sides in the center of the junction. In the second, less
easily achievable, case, sketched in Fig.1(b), the control
line is perpendicular to the junction plane and goes through a small
hole drilled in the substrate in between the gap of a ring shaped
JTJ. In the former case the bias current density is uniform, while
in the latter case the current distribution can be expressed in
terms of a $\delta$-function\cite{PRB10}. It is quite evident that
ABC are achieved in both electrode configurations.

\begin{figure}[ht]
\centering
\subfigure[]{\includegraphics[width=6cm]{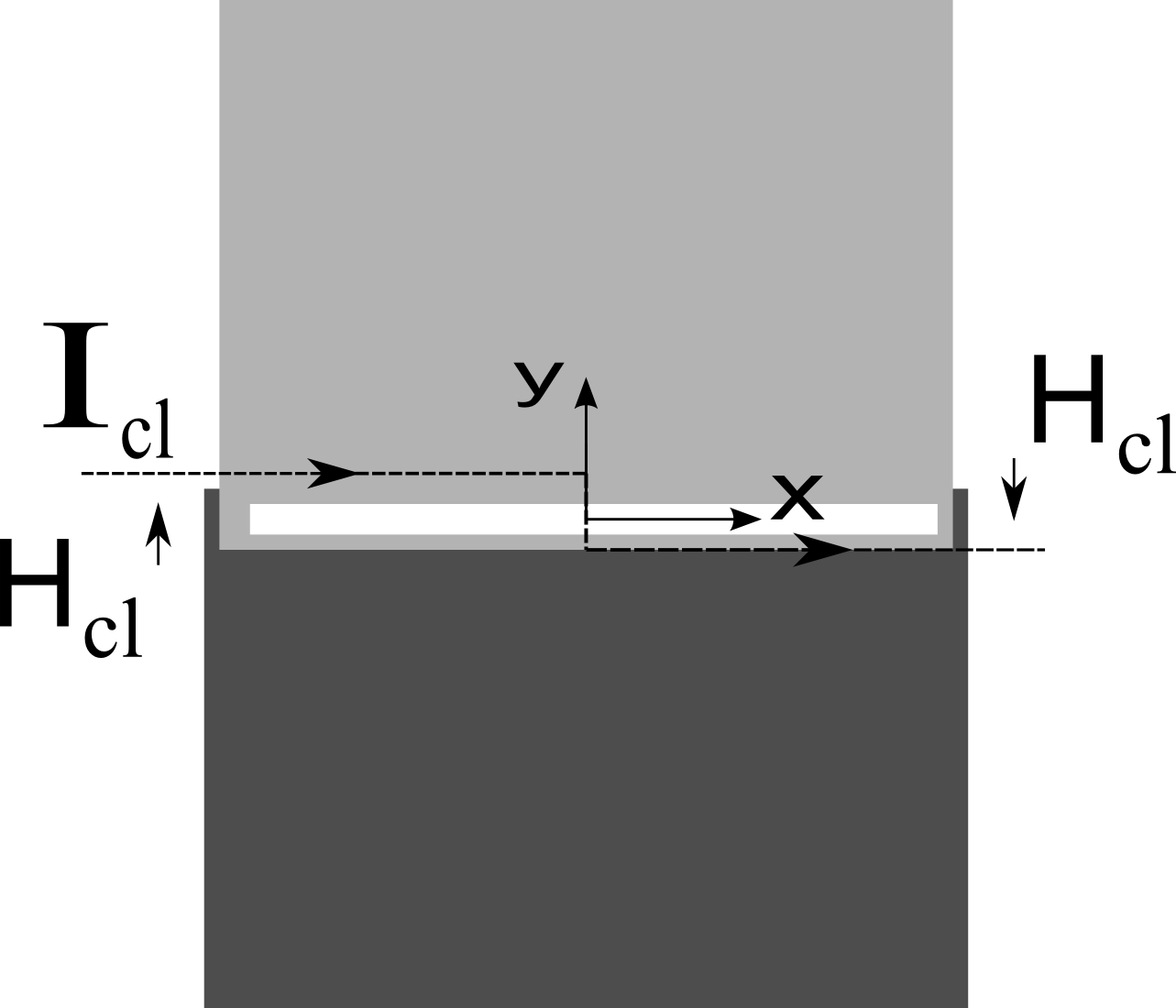}}
\subfigure[]{\includegraphics[width=4cm]{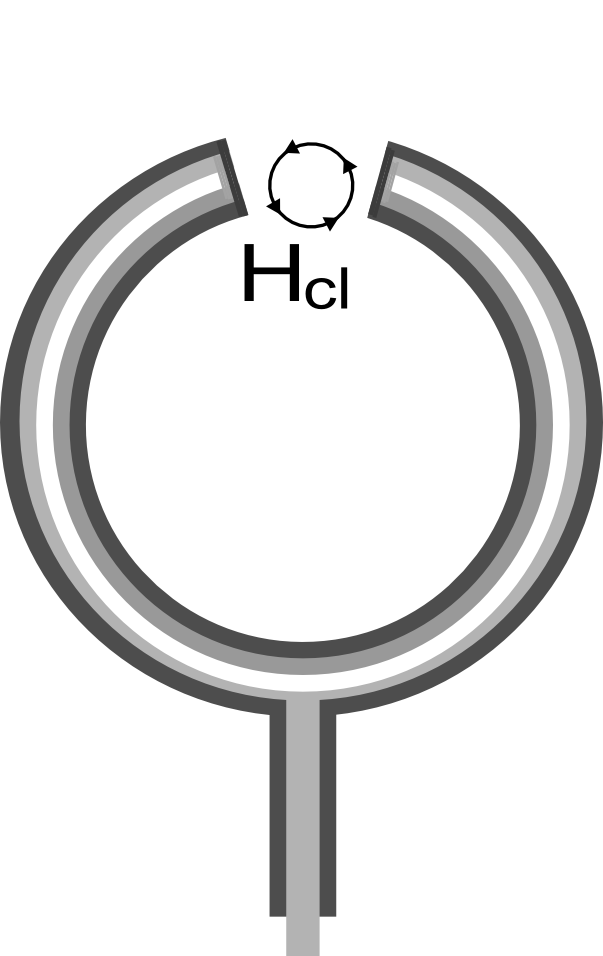}}
\caption{Sketches of planar Josephson tunnel junctions in the asymmetric magnetic field $H_{cl}$ generated by properly designed control lines. (a) (uniformly biased) linear and (b) ($\delta$-biased) gapped annular Josephson tunnel junctions. The base electrode is in dark gray, the top electrode is in light gray and the junction area is white.}
\label{2D}
\end{figure}

\section{ASYMMETRIC BOUNDARY CONDITIONS}
\noindent The ABC have several important implications. We begin by
computing the Josephson current $i_j$ carried by the junction. From
Eq.(\ref{PDE}) we easily get:

\begin{equation}
i_j=  \frac{1}{l} \left[ \phi_x(l/2)- \phi_x(-l/2) \right] + i_b=\frac{h_R-h_L}{l} +i_b.  \label{ij}
\end{equation}

\noindent With the classical boundary conditions, the first term of
the right side of Eq.(\ref{ij}) vanishes, so that $i_j=i_b$.
However, with ABC we get:

\begin{equation}
i_j=  \frac{2h_{a}}{l}+i_b,  \label{ijj}
\end{equation}

\noindent i.e., the zero-voltage current also depends on the
amplitude of the asymmetric magnetic field. More precisely, the
applied magnetic field acts as an extra field-dependent d.c. current
source $i_h=2h_{a}/l$ in parallel to the external current source
$i_b$. In real units the normalized current $i_h$ corresponds to a
supercurrent $I_H= i_h J_c WL= 2H_RW$ proportional to the junction
width $W$ (e.g., with $H_R = 10$A/m and $W=3\mu$m, then
$I_H=60\mu$A). Eq.(\ref{ij}) suggests that a planar JTJ can be used
as a magnetic first order gradiometer based on the readout of the
current $i_h$ supplied by a stand-alone (unbiased) JTJ. However, it
should be realized that, since $i_h$ is a zero-voltage current, the
internal impedance of the current source is null, so that no power
can be delivered to an external resistive load.

\noindent From a mathematical point of view, the peculiarity of the
ABC is that, provided that the current distribution is symmetric
$\gamma(-x)=\gamma(x)$, the solutions of Eq.(\ref{PDE}) have to be
even functions, $\phi(x)=\phi(-x)$, meaning that the phase
difference between the two junction extremities is always null:
$\phi(l/2)- \phi(-l/2) = 0$. As a consequence of the phase parity,
we could limit our analysis to the spatial range $[0, l/2]$, with
the condition in the origin to be found as follows. From
Eq.(\ref{PDE}), for any $0<x_0<l/2$, we can write:

\begin{equation}
\phi_x(x_0) - \phi_x(-x_0)  =   \int_{-x_0}^{x_0}  \sin \phi(x) dx - \int_{-x_0}^{x_0}  \gamma(x) dx.  \label{jump0}
\end{equation}

\noindent Taking the limit $x_0 \to 0$, the first integral vanishes,
since $\phi(x)$ is a continuous function. If also the bias current
density $\gamma(x)$ is continuous, then the second integral vanishes
too, enforcing $\phi_x(0+)- \phi_x(0-) =0 $. Being the phase
derivative an odd function, $\phi_x(0+)+ \phi_x(0-) =0 $, it
vanishes in the origin:

\begin{equation}
\phi_x(0)=0.  \label{bc0}
\end{equation}

\noindent However, this is no longer true when the bias profile is
discontinuous. In the case of a $\delta$-shaped bias,
Eq.(\ref{jump0}) leads to a discontinuity of the phase gradient in
the origin\cite{kuprianov,PRB10}: $\phi_x(0+)-
\phi_x(0-)=\gamma_{\delta}$. Exploiting again the symmetry property
of the phase gradient, we have: $\phi_x(0-) = \gamma_{\delta}/2$ and

\begin{equation}
\phi_x(0+) = -\gamma_{\delta}/2.
\label{bc1}
\end{equation}

\noindent To summarize, an overlap junction with ABC obeys
Eq.(\ref{PDE}) with the condition at $x=l/2$ as in Eq.(\ref{bc}):

\begin{equation}
\phi_x(l/2) = h_{a},
\label{bcl}
\end{equation}

\noindent and the condition in the origin given by Eq.(\ref{bc0}) or
(\ref{bc1}) in the cases of continuous or $\delta$-shaped bias,
respectively.

\section{SMALL JTJs}

\noindent Since the phase profile of a junction with ABC has to be
even, then, for small junctions ($l<1$), we can use the trial
function $\phi(x)=ax^2+b|x|+ \phi_0 $, with the parameters $a$ and
$b$ to be determined from the boundary conditions and $\phi_0$
treated as an integration constant. We will consider separately the
cases of uniform bias and that of a $\delta$-shaped bias
profile\cite{note}. In the former case, in order to fulfill the
condition in (\ref{bc0}), then $b=0$ and, to satisfy (\ref{bcl}),
$a=h_{a}/l$, so that: $ \phi(x)=h_{a} x^2/l+\phi_0$. The Josephson
current $i_j$ can be computed from the above quadratic expression:

\begin{equation}
i_j(h_{a})= \sqrt{\frac{2\pi}{h_{a}l}} \left[ S \left( \sqrt{\frac{h_{a}l}{2\pi}} \right) \cos \phi_0 +  C \left( \sqrt{\frac{h_{a}l}{2\pi}} \right) \sin \phi_0 \right],
\label{ij3}
\end{equation}


\noindent in which we have introduced Fresnel's integrals defined
by: $ \int_{0}^{\bar x} \sin a x^2 dx =\sqrt{{\pi}/{2a}}\, S \left(
\sqrt{{2a}/{\pi}}\,{\bar x} \right)$ and $ \int_{0}^{\bar x} \cos a
x^2 dx =\sqrt{{\pi}/{2a}}\, C \left( \sqrt{{2a}/{\pi}}\,{\bar x}
\right)$ ($a>0$). The critical current $i_c$ can be found by maximizing
(\ref{ij3}) with respect to $\phi_0$. Introducing the quantity
$h_e=\sqrt {|h_{a}|l/2\pi}$, we get a Fresnel (or near-field)
magnetic diffraction pattern:

\begin{equation}
i_c(h_e)= \frac{\sqrt{ S^2(h_e)+ C^2(h_e) }}{h_e},
\label{ic3}
\end{equation}

\noindent with $\phi_0(h_e)=\tan^{-1} C(h_e)/S(h_e)$. In the limit
$h_e \to 0$, $S(h_e)\approx 0$ and $C(h_e)\approx h_e$ so that
$i_c(0)=1$. In the opposite limit, i.e., for $h_e \to \infty$,
$S(h_e)\approx C(h_e) \approx 1/2$, so that $i_c(h_{a})\approx\sqrt
{\pi/h_{a}l}$. Fig.2 shows the dependence of the
critical current $i_c$ on the product $h_a l$ (that is independent on $\lambda_j$).

\noindent In the case of $\delta$-shaped bias, then, in order to
satisfy the conditions in  (\ref{bc1}) and (\ref{bcl}),
$b=-\gamma_{\delta}/2$ and $2a=(2h_{a}+\gamma_{\delta})/l=i_j $, so
that: $\phi(x)= i_j x^2/2-\gamma_{\delta}|x|/2+ \phi_0$. When the
modulus of the ratio $li_j/\gamma_{\delta}$ is much smaller than unity, the
quadratic term can be disregarded. $i_j = 0$ means that
$\gamma_{\delta} =-2h_{a}$, i.e., the solution is piecewise linear:

\begin{equation}
\phi(x)=h_{a} \left|x\right|+\phi_0. \label{db}
\end{equation}


\noindent  Eq.(\ref{db}) works well either for small junctions
($l<1$) or near the pattern minima ($i_j\approx 0$) or for large
field values ($h>>2$). The Josephson current corresponding to
Eq.(\ref{db}) can be computed:

\begin{equation}
i_j(h_{a})= \frac{2}{l} \left[\cos \phi_0 \left( 1-\cos \frac{h_{a}l}{2} \right)+ \sin \phi_0 \sin \frac{h_{a}l}{2} \right].
\label{ij4}
\end{equation}

\noindent The critical current $i_c(h_{a})$ can be found
by maximizing (\ref{ij4}) with respect to $\phi_0$, to give:


\begin{equation}
i_c(h_{a})=  \frac{\sin h_{a}l/4}{h_{a}l/4}.
\label{ic4}
\end{equation}

\noindent The integration constant is the sawtooth function:

\begin{equation}
\phi_0(h_{a})=\tan^{-1} \textrm{cotan}\, \frac{h_{a}l}{4}. \label{phidb}
\end{equation}

\noindent In other words, for a small $\delta$-biased JTJ with ABC
we expect a Fraunhofer-like magnetic diffraction pattern with a
field-periodicity twice larger than that of a small uniformly biased
 JTJ in a uniform magnetic field $h^{u}$, for which $i_c(h^{u})=\sin
(h^{u}l/2) /(h^{u}l/2)$.

\begin{figure}[tb]
\centering
{\includegraphics[width=6cm]{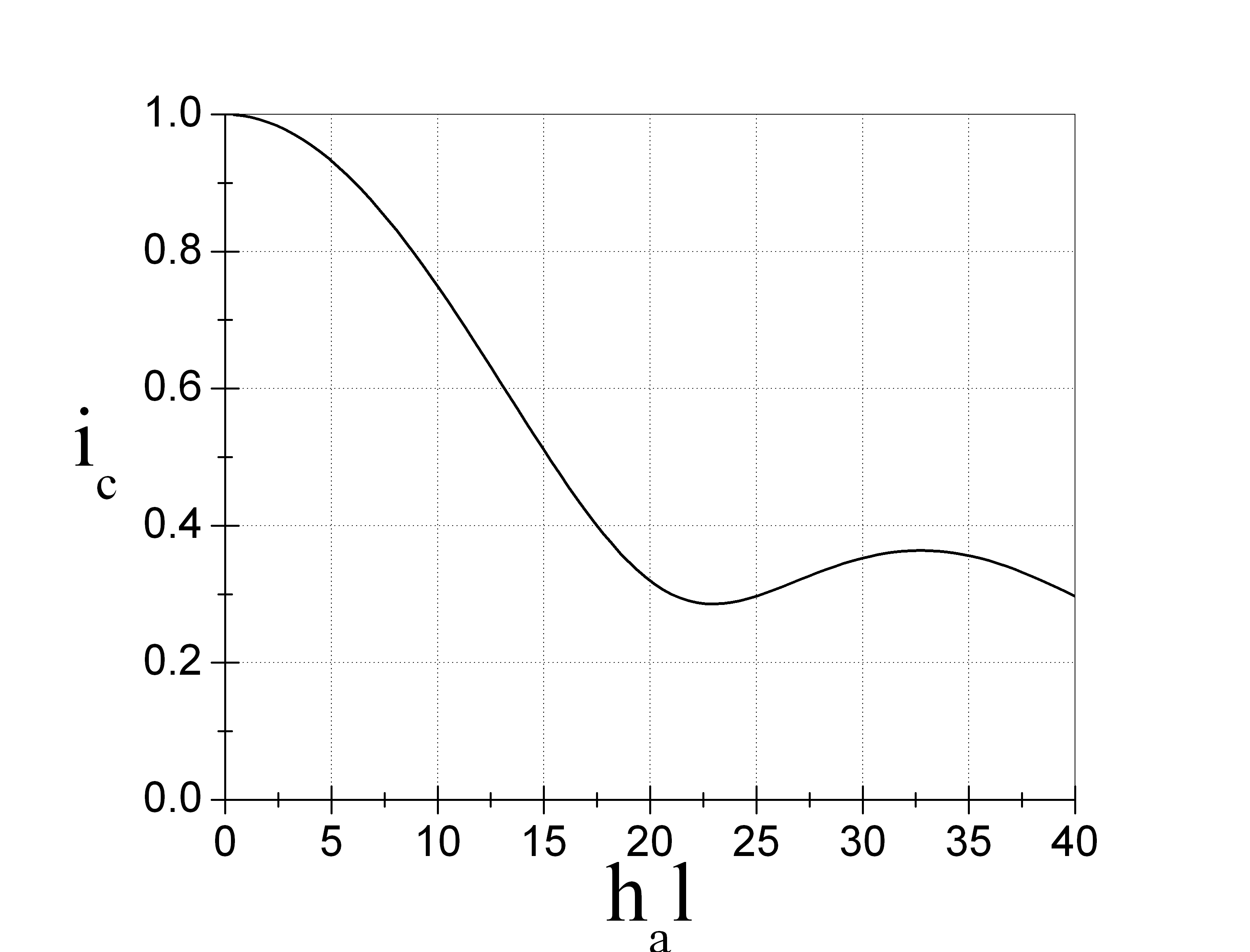}}
\caption{Fresnel-like magnetic diffraction pattern $i_c(|h_{a}|l)$ as in Eq.(\ref{ic3})in which $h_e=\sqrt {|h_{a}|l/2\pi}$.}
\label{fresnel}
\end{figure}

\section{LONG JTJs}
\noindent Eq.(\ref{PDE}) with ABC has been numerically integrated in
order to find the magnetic diffraction patterns $i_c(h_{a})$ of JTJs
having lengths larger that $\lambda_j$. The details of the numerical
technique can be found in Ref.\cite{PRB10} together with the
analogous results in the case of symmetric boundary conditions.
Figs.3(a)-(d) show the numerically obtained $i_c(h_{a})$
for JTJs with ABC having normalized lengths $l=2,4,8$ and $16$,
respectively: the full dots refer to the uniform bias, while the
open dots correspond to the point injected current. Note the very
good agreement between the theoretical dependence of
Fig.2 and the result of the numerical simulations for
$l=2$ shown by the dashed line of Fig.3(a).

\noindent The validity of Eq.(\ref{ijj}) has been numerically checked both in the
case of uniform bias and $\delta$-shaped bias. In the latter case,
for $l=2$, the phase profile could be well approximated by $\phi(x)=h_{a}
x^2/l+\phi_0$ for $|h_{a}| <1$ ($0<\phi_0<\pi/2$) and by
Eq.(\ref{db}) for $|h_{a}|>1$ [with $\phi_0$ given by
Eq.(\ref{phidb})]. For $l \geq 4$ the linear approximation is good
for $|h_{a}|>2-3$. Generally speaking, it is observed that the
magnetic diffraction patterns become more and more asymmetric as
the junction normalized length increases. Parenthetically, we note
that what is measured in the experiments is the maximum bias current
that by virtue of Eq.(\ref{ij}) can be quite different from the
critical current. This might explain why the effects of a
non-perfectly uniform magnetic field have never been reported in the
literature so far.

\noindent An asymmetric magnetic field drastically modifies also the
dynamics of a planar JTJ. It is easy to see that the Josephson phase
parity does not allow the standing wave resonances leading to
current singularities such as Fiske and flux flow steps observed in
presence of a uniform external field\cite{kukik}. In fact, the
magnetic field with opposite directions at the junction ends forces
a chain of fluxons entering on one side and a chain of antifluxons
entering on the other side, with the total magnetic flux in the
barrier being always null. Furthermore, a small amplitude asymmetric
field provides an asymmetric tuning of the average speed of one (or
more) soliton shuttling back and forth along the junction, opposite
to the symmetric tuning typical of a uniform field\cite{levring}.

\begin{figure}[ht]
\centering
\subfigure[ ]{\includegraphics[width=6cm]{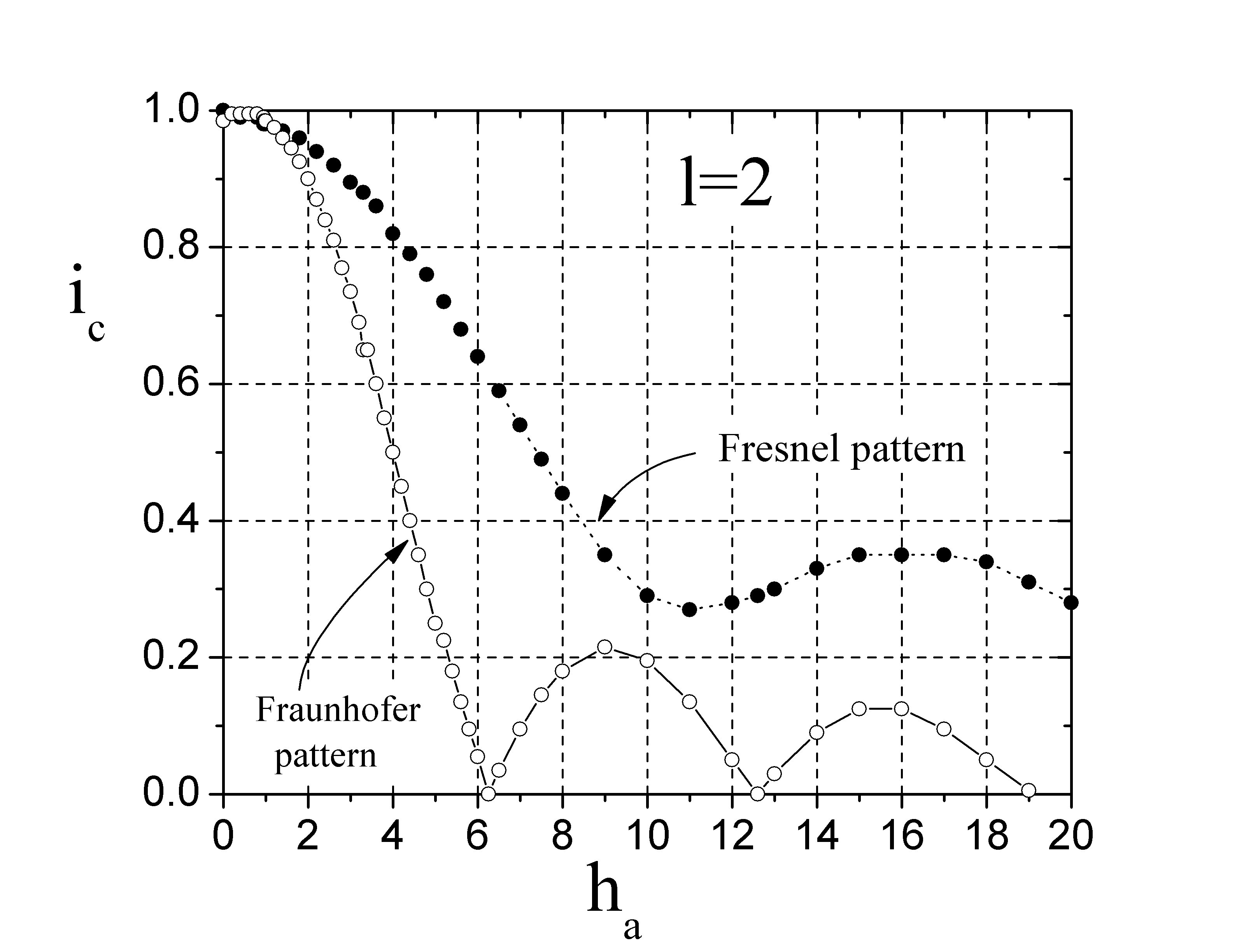}}
\subfigure[ ]{\includegraphics[width=6cm]{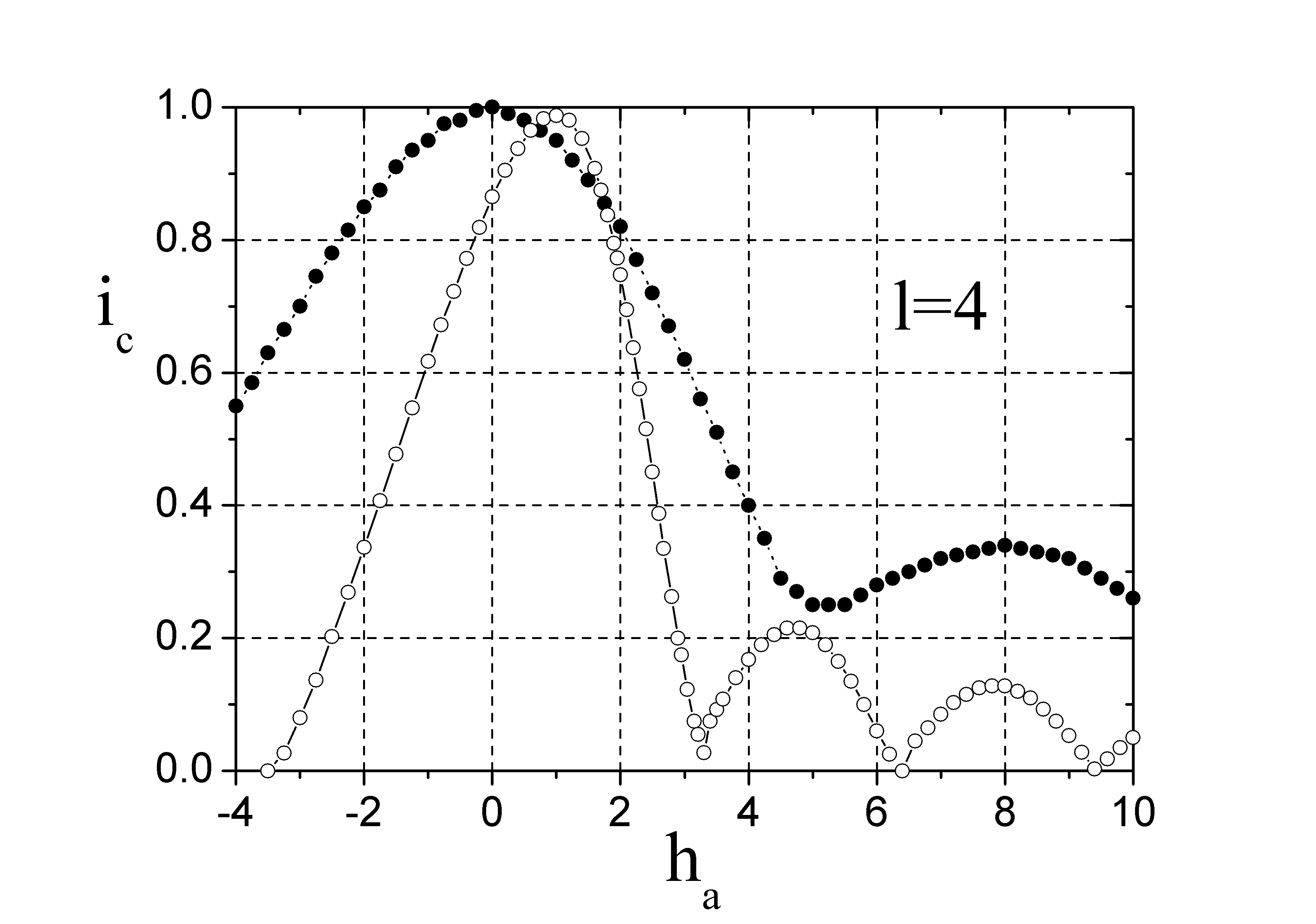}}
\subfigure[ ]{\includegraphics[width=6cm]{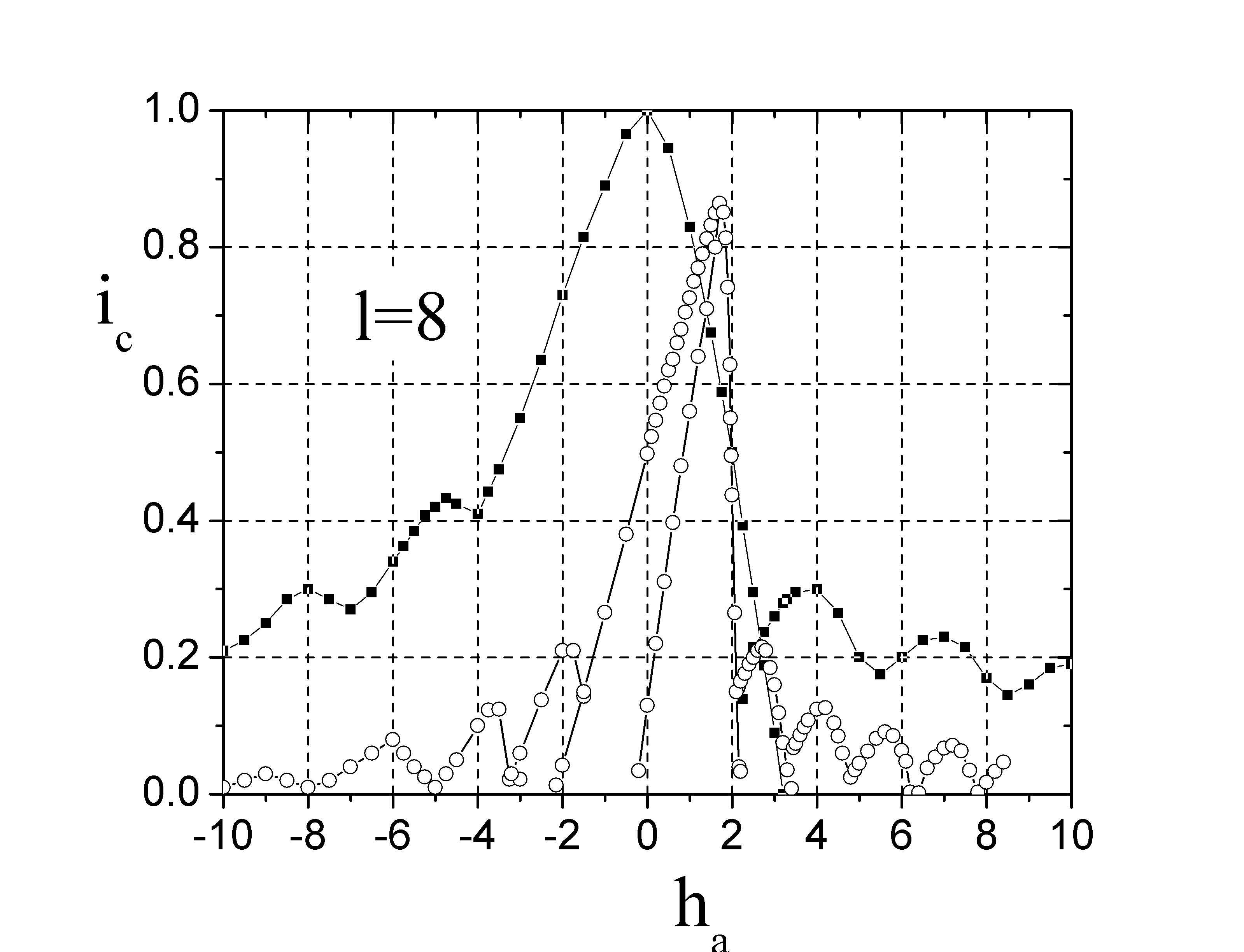}}
\subfigure[ ]{\includegraphics[width=6cm]{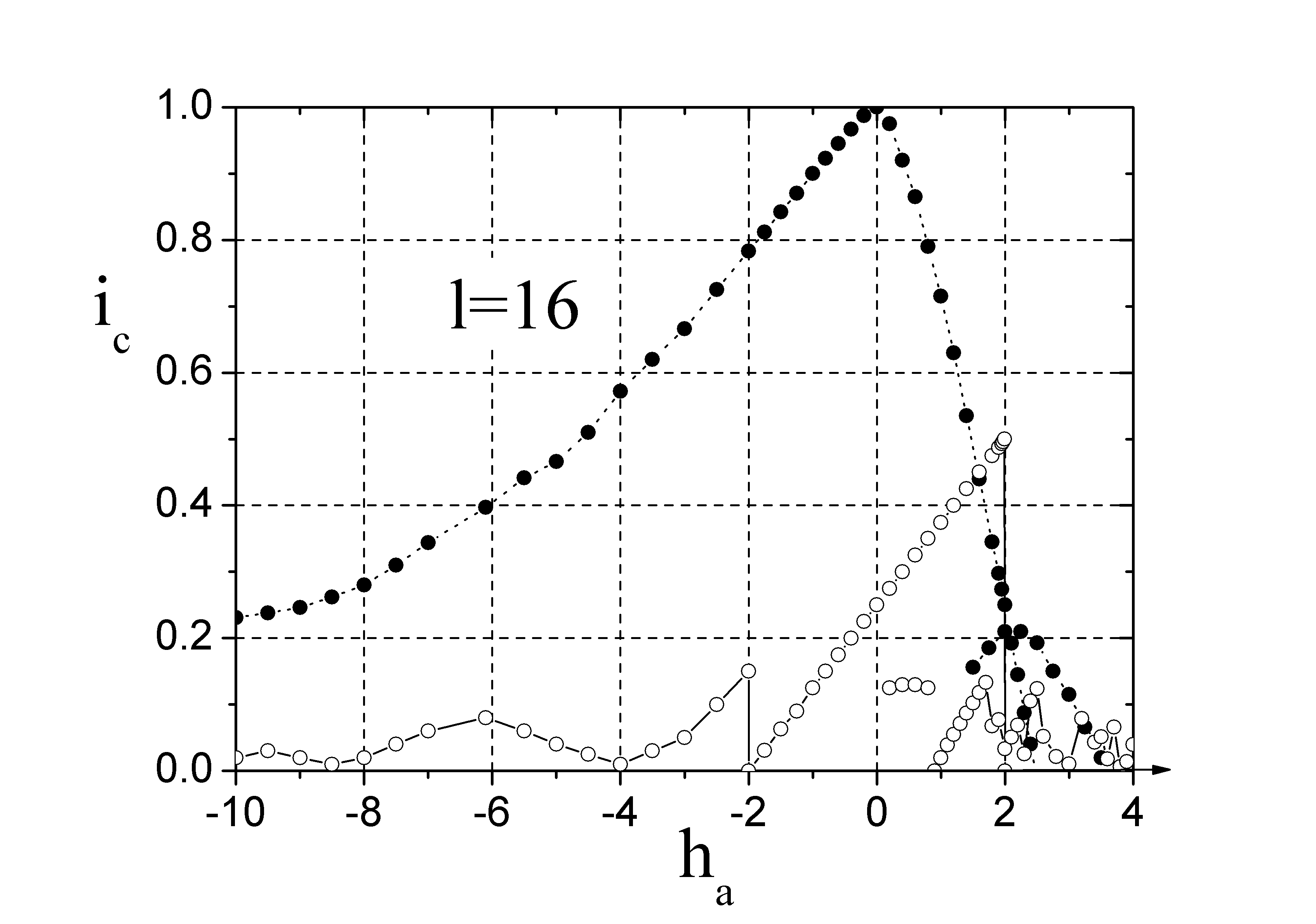}}
\caption{Numerically computed magnetic diffraction pattern for intermediate length and long Josephson tunnel junctions with asymmetric boundary conditions: $i_c$ vs. $h_{a}$ for $l=2,4,8,16$. The full dots refer to uniformly biased junctions, while the open circles correspond to $\delta$-biased ones.}
\label{asymm}
\end{figure}

\section{CONCLUDING REMARKS}
\noindent In summary, it has been discussed how the magnetic
properties of a planar Josephson tunnel junction change when the
textbook assumption of a perfectly homogeneous magnetic field is reverted
into a fully asymmetric one. In the most general case when the
magnetic field has different amplitudes at the junction extremities,
due to the system non-linear nature, the problem cannot be split in
two subproblems with properly chosen symmetric and asymmetric
boundary conditions, unless the conditions for linearizing the
current-phase relationship occur\cite{vaglio}. The experimental
verification of our findings has been planned.

\section*{Acknowledgements}
\noindent The author thanks  V.P. Koshelets, J. Mygind, and R.J.
Rivers for useful and stimulating discussions.



\begin{thebibliography}
\bibitem{note} When the junction normalized length is much smaller than unity, then any electrode configuration will result in an uniform bias.
\end{thebibliography}
\end{document}